\documentstyle[aps,prb,graphicx]{revtex}

\begin{document}

\twocolumn[\hsize\textwidth\columnwidth\hsize\csname@twocolumnfalse\endcsname
\author{Donavan Hall$^{1}$, E. Palm$^{1}$, T.
Murphy$^{1}$, S. Tozer$^{1}$, Z. Fisk$^{1}$,
U. Alver$^{2}$, R. G. Goodrich$^{2}$, J. L. Sarrao$^{3}$, P.G. 
Pagliuso$^{3}$,and Takao Ebihra$^{4}$}
\address  {$^{1}${\small \ National High Magnetic Field
Laboratory, Florida State University, Tallahassee, FL 32310.}}
\address{ $^{2}${\small Department of Physics and Astronomy,
Louisiana State University, Baton Rouge, LA 70802.}}
\address {$^{3}$ {\small Los Alamos National Laboratory, MST-10, Los
Alamos, NM 87545,}}
\address {$^{4}$ {\small Department of Physics, Faculty of Science, Shizuoka University, 836 Ohya, Shizuoka 422-8529, Japan}}
\title{The Fermi surface of CeCoIn$_{5}$: dHvA.}
\date{February 27, 2001}
\maketitle
\
\begin{abstract}
Measurements of the de Haas - van Alphen effect in the normal state of the
heavy Fermion superconductor CeCoIn$_{5}$ have been carried out using a
torque cantilever at temperatures ranging from 20 to 500 mK and in fields
up to 18 tesla.  Angular dependent measurements of the extremal Fermi
surface areas reveal a more extreme two dimensional sheet than is found
in either CeRhIn$_{5}$ or CeIrIn$_{5}$. The effective masses of the measured
frequencies range from 9 to 20 m*/m$_{0}$.
\\
\end{abstract}
\maketitle
\\
]

\section{Introduction}

The compounds CeMIn$_{5}$ (M=Co, Ir, Rh) are a fascinating family of heavy
fermion superconductors \cite{Hegger2000,Thompson2000}. These materials
crystallize in the tetragonal HoCoGa$_{5}$ structure and are built of
alternating stacks of CeIn$_{3}$ and MIn$_{2}$. The compounds CeCoIn$_{5}$
and CeIrIn$_{5}$ have superconducting transition temperatures of 2.3 K and
0.4 K, respectively, whereas CeRhIn$_{5}$ orders antiferromagnetically at
3.8 K at ambient pressure, but applied pressures of order 16 kbar can induce
an apparently first-order transition from the magnetically ordered state to
a superconducting one with T$_{c}$= 2.1 K. Below 1.4 K the transition from
the normal state to the superconducting state as a function of magnetic
field in CeCoIn$_{5}$ has been found to be first order at ambient pressure%
\cite{Murphy}. The particular attraction of these materials, then, is that
not only do they represent a substantial increase in the number of known
heavy fermion superconductors but also they appear to be quasi-2D variants
of CeIn$_{3}$, an ambient-pressure antiferromagnet in which
superconductivity can be induced at 25 kbar and 200 mK \cite{GGL-Nature} and
have properties of unconventional superconductivity. Movshovich et al.\cite
{Movshovich2001} have shown a power law temperature dependence in the
specific heat and thermal conductivity suggesting strongly that the
super-conductivity in CeCoIn$_{5}$ is unconventional. If one can demonstrate
that the reduced dimensionality is responsible for the factor of 10 increase
in superconducting T$_{c}$, the impact of these compounds on the physics of
superconductivity will far exceed that of just another family of heavy
fermion superconductors.

As a first step in determining the dimensionality and nature of the
electronic structure in CeCoIn$_{5}$ we report the results of de Haas - van
Alphen (dHvA) measurements that show the quasi-2D nature of the FS as well
as the large effective masses of electrons on this FS. We also compare our
results to previously reported dHvA studies of CeRhIn$_{5}$\cite{Hall} and
CeIrIn$_{5}$\cite{Haga2000}.

\section{Measurements}

The measurements reported here were performed at the National High Magnetic
Field Laboratory, Tallahassee, FL using a rotating torque cantilever
magnetometer designed for operation at low temperatures between 20 and 500
mK in applied fields ranging from 5 to 18 T. Complete field rotations in the
[100] and [001] planes of the tetragonal structure were made, and
temperature dependent measurements of the dHvA amplitudes were studied.

The sample was grown from an In flux\cite{Petrovic2000} and etched in a 25\%
HCl in H$_{2}$O solution down to a small plate that was mounted on the
cantilever with GE varnish. The sample was mounted in multiple orientations
with respect to the cantilever to remove any possibility of systematic
instrumental error which might affect frequency and mass determinations.

In the dHvA measurements the oscillatory part of the magnetization of the
sample is measured as a function of field. The resulting signal is periodic
in 1/B. This oscillatory magnetization $\widetilde{M}$ is given by the
Lifshitz-Kosevitch (LK) equation (see Ref. \onlinecite{Shoenberg} for the
mathematical details):

\begin{eqnarray}
\widetilde{M} &=&-2.602\times 10^{-6}\left( \frac{2\pi }{HA^{\prime \prime }}
\right) ^{1/2}  \nonumber \\
&&\times \frac{\Gamma FT\exp (-\alpha px/H)}{p^{3/2}\sinh (-\alpha pT/H)}
\sin \left[ \left( \frac{2\pi pF}{H}\right) -\frac{1}{2}\pm \frac{\pi }{4}%
\right] ,  \label{eq:LK}
\end{eqnarray}

\noindent where $\alpha $ = 1.47(m/m$_{0}$) $\times $ 105 G/K, $A''$ is the
second derivative of the area of the Fermi surface (FS) cross-section that
is perpendicular to the applied field, $\Gamma $ is the spin reduction
factor, $p$ is the harmonic number, and $x$ is the Dingle temperature. The
frequencies of the dHvA oscillations are proportional to extremal areas of
the FS. Applying this formula to extract the FS properties of a heavy
fermion material presents no new complications\cite{Wasserman}.

The measured signal from the torque cantilever is a voltage
proportional to the gap between the flexible cantilever plate to which
the sample is attached and a fixed (reference) plate.  The total gap
separation is measured as a capacitance using a precision capacitance
bridge that can detect changes to better than a part in 10$^{6}$.  The
measured oscillations in the torque arise from anisotropy in the Fermi
surface, such that

\begin{equation}
\widetilde{\tau }=\frac{-1}{F}\frac{dF}{d\Theta }\widetilde{M}HV
\label{eq:LKtorque}
\end{equation}

\noindent where F is the dHvA frequency, $\Theta $, is the angle of the
applied field, $\widetilde{M}$ is the LK expression above, and V is the
volume of the sample. It should be noted that for field directions near high
symmetry axes where extremal FS areas go through maxima or minima such that $%
\frac{dF}{d\Theta }$ becomes small, the signals measured with a torque
cantilever also become small.

\section{Results}

\begin{center}
\begin{figure}[tbp]
	\includegraphics[scale=0.45]{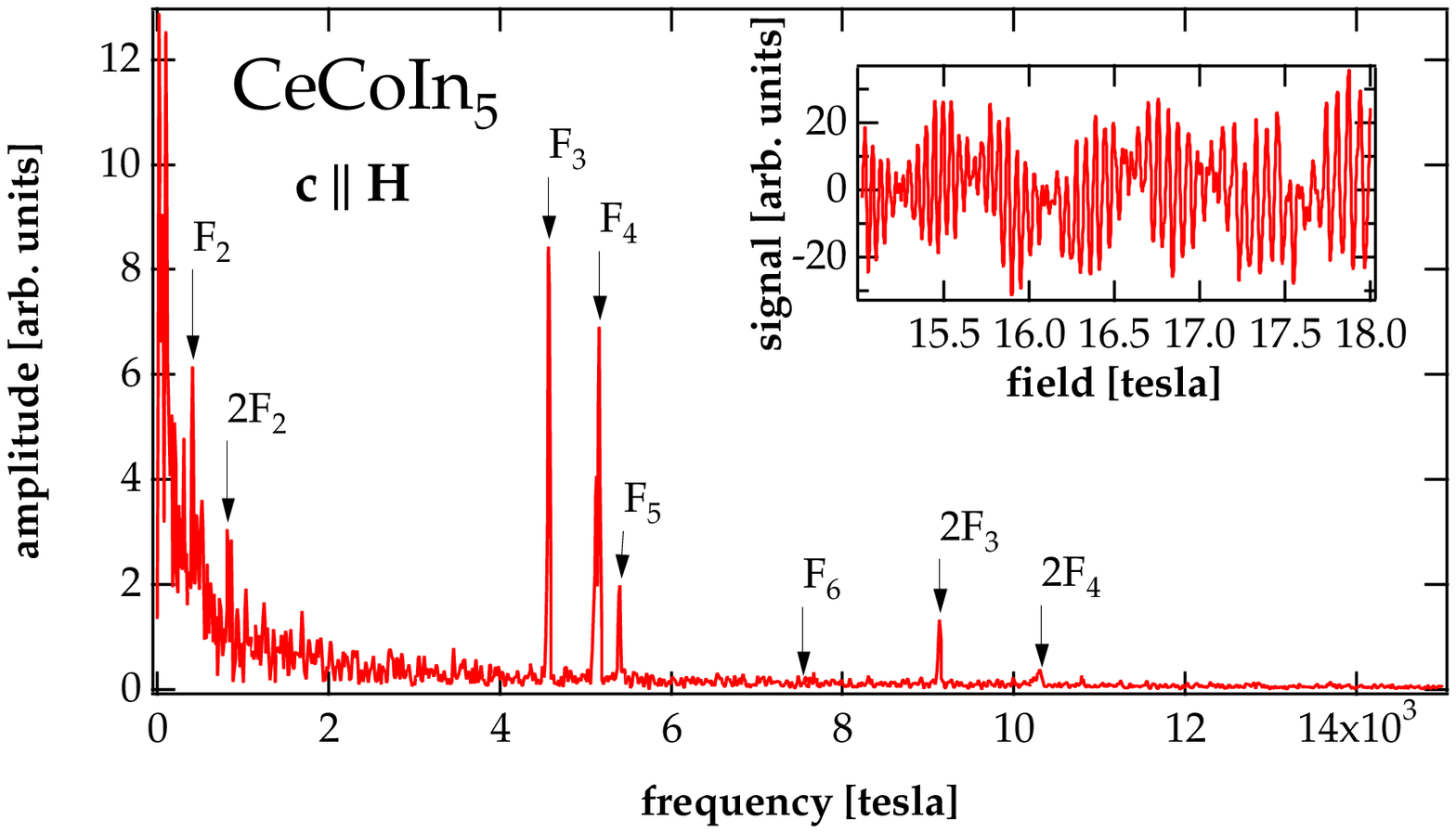}
	\caption{}
\label{lowfft}
\end{figure}
\end{center}

For the field directed along the [001] axis oscillations are clearly seen at
the lowest temperatures (20 mK) as shown in Fig. \ref{lowfft} along with an FFT of the
data. The frequencies of these oscillations are listed in Table 
\ref{freq001}. As the
field is rotated away from the [001] the signals become stronger due to the
larger values of $\frac{dF}{d\Theta }$ away from the axis. The determination
of the dHvA frequencies for angles between [001] and [100] and between [001]
and [110] were possible with the results displayed in Figure 
\ref{rotations}. The masses for the field applied along the [111]
where $\frac{dF}{d\Theta }$ is large were measured and had values from
nearly 9 to 20 m*/m$_{0}$.  The
effective masses for carriers on three orbits for the field applied along
the [111] were determined, the results of which are summarized in Table 
\ref{cmass111}.
Mass determinations for the [100] and [001] principles axes were not
possible in these measurements due to a near zero slope in the frequency vs.
angle plots that causes very small signals and a rapidly diminishing signal
with increasing temperature. This temperature behavior is the signature for
high effective masses as the high masses in Table \ref{cmass111} 
illustrate. 

\section{Discussion}

The dHvA measurements on CeIrIn$_{5}$ \cite{Haga2000} and CeRhIn$_{5}$ \cite
{Hall} find multiple branches for rotations in the [001] -[100] and [001] -
[110] planes of the tetragonal structure. Most of these branches are
associated with large quasi-2D undulating cylinders that show the expected
1/cos($\theta $) dependence with $\theta $ being the angle at which the
field is applied away from the [001] axis. Band structure calculations, both
in Ref. \onlinecite{Haga2000} and in Ref. \onlinecite{Hall} show, in
addition, that several small pieces of FS should exist in both CeRhIn$_{5}$
and CeIrIn$_{5}$.

We find a very similar situation in CeCoIn$_{5}$ in that the F$_{3}$, F$_{4}$%
, and F$_{5}$ branches in Figure 2 are closely spaced in frequency,
corresponding to extremal areas on an open 2D undulating cylinder extending
along the [001] direction. However, fewer frequencies are observed for CeCoIn%
$_{5}$ than was the case for CeRhIn$_{5}$\cite{Hall}or CeIrIn$_{5}$\cite
{Haga2000}. Some of the frequencies observed in these two cases are
attributed to holes in the cylinders and the band structure calculations
bear out this assignment. Because the mass enhancements expected in all of
these materials is comparable, based on heat capacity data\cite{Petrovic2001},
the reduced number of observed frequencies seen here cannot be attributed
to particularly heavy carriers in CeCoIn$_{5}$. The cylindrical surface in
CeCoIn$_{5}$ appears to be closed with no holes, and much more 2D like than
in either of the other two cases. The fact that we observe three closely
spaced frequencies for the field applied along [001] indicates that there
are small undulations on this cylinder giving rise to the three extremal
areas. Based on comparisons of the observed closely spaced frequencies near
6000 T, the magnitude of this undulation in CeCoIn$_{5}$ is 
approximately 50\% less than that
observed in CeRhIn$_{5}$ and CeIrIn$_{5}$. In addition, the number of low
frequencies in CeCoIn$_{5}$ is much smaller than in either the Rh or Ir
analogs so there are a smaller number of electrons exhibiting 3D behavior.

These results indicate that many of the transport properties and cooperative
phenomena that are seen in CeCoIn$_{5}$ should be much more 2D in character
than those found in either CeRhIn$_{5}$ or CeIrIn$_{5}$. The fact that the T$%
_{c}$ in CeCoIn$_{5}$ is 5 times higher than that observed in CeIrIn$_{5}$
would suggest that the increasingly two-dimensional electronic structure has
a direct correlation with enhanced T$_{c}$. Studies of the Fermi surface of
CeRhIn$_{5}$ at pressures adequate to produce superconductivity would be
valuable in confirming this supposition because the T$_{c}$ of CeRhIn$_{5}$
under pressure is comparable to that of CeCoIn$_{5}$.

\section{Acknowledgments}

This work was supported in part by the National Science Foundation under
Grant No. DMR-9971348 (Z. F.). A portion of this work was performed at the
National High Magnetic Field Laboratory, which is supported by NSF
Cooperative Agreement No. DMR-9527035 and by the State of Florida. Work at
Los Alamos was performed under the auspices of the U. S. Dept. of Energy.

\begin{center}
FIGURE CAPTIONS
\end{center}

Figure \ref{freq001}. The Fourier spectrum of the dHvA oscillations shown in the inset
for the applied field along [001]. Six fundamental frequencies and
associated harmonics are observed.

\begin{center}
\begin{figure}[tbp]
	\includegraphics[scale=0.5]{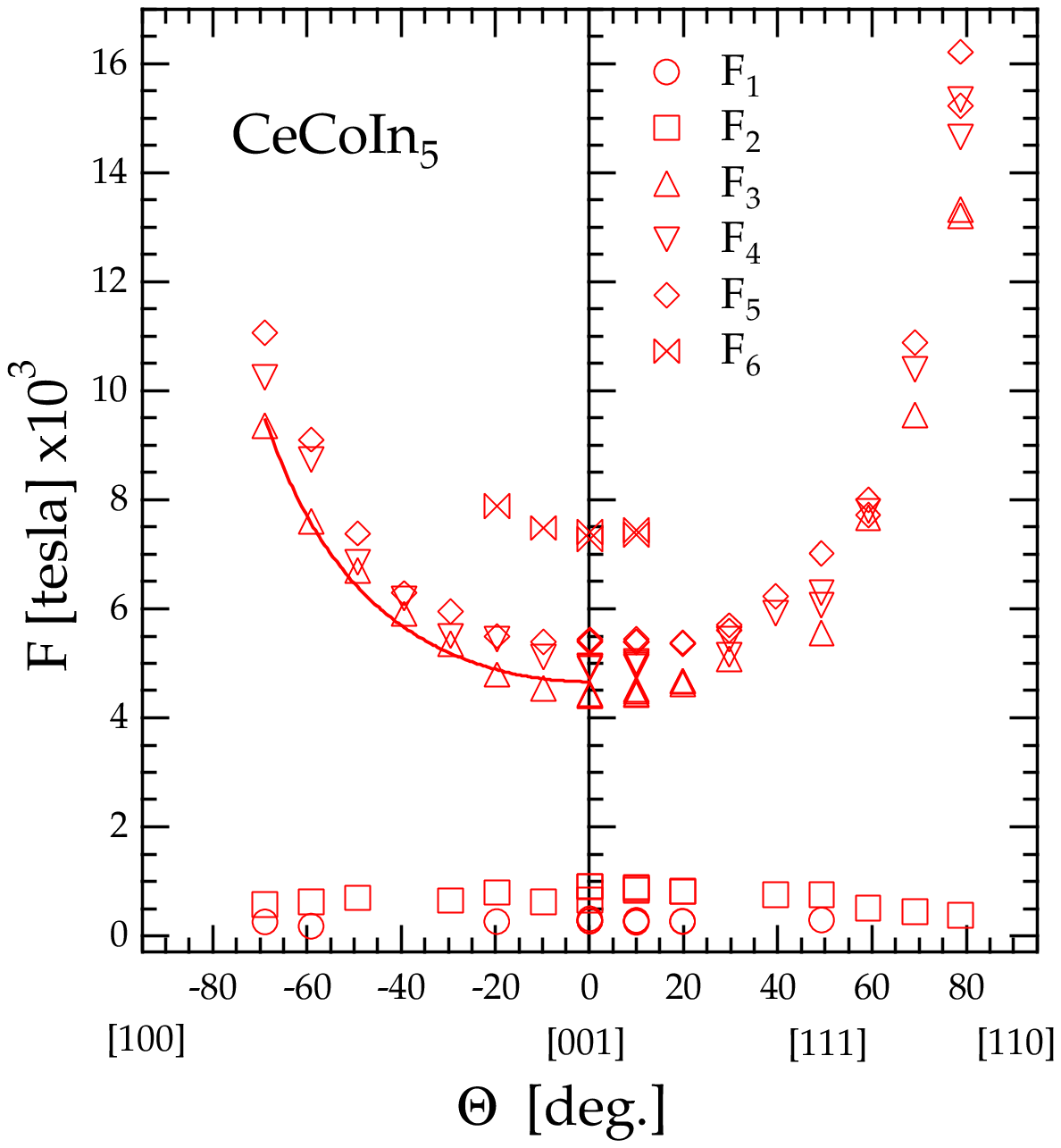}
	\caption{}
\label{rotations}
\end{figure}
\end{center}

Figure \ref{rotations}. The angular dependence of the dHvA frequencies for CeCoIn$_{5}$ is
shown here for rotations around [001]. The solid line shows a fit to the
expected 1/cos$\Theta $ dependence for a 2D FS. The low frequencies are due
to small 3D ellipsoidal pockets.

\begin{table}[tbp]
\caption{Measured dHvA frequencies for CeCoIn$_{5}$ with $B$ along [001].
Both F$_{1}$ and F$_{6}$ are difficult to see in this FFT. The amplitude of F%
$_{1}$ grows with increasing angle. F$_{6}$ is also more evident at the
higher angles, but it observable in this data set if the high field data is
analyzed apart from the low field part of the sweep.}
\label{freq001}
\begin{center}
\begin{tabular}[t]{cc}
Symbol & F (T) \\ \hline
F$_{6}$ & 7535 \\
F$_{5}$ & 5401 \\
F$_{4}$ & 5161 \\
F$_{3}$ & 4566 \\
F$_{2}$ & 411 \\
F$_{1}$ & 267
\end{tabular}
\end{center}
\end{table}

\begin{table}[tbp]
\caption{Measured masses for dHvA frequencies for CeCoIn$_{5}$ with $B$
along [111].}
\label{cmass111}
\begin{center}
\begin{tabular}[t]{ccc}
Symbol & F (T) & m*/m$_{0}$ \\ \hline
F$_{4}$ & 6064 & 20.3 $\pm $ 0.7 \\
F$_{3}$ & 5550 & 11.2 $\pm $ 0.2 \\
F$_{2}$ & 760 & 8.7 $\pm $ 2.2
\end{tabular}
\end{center}
\end{table}

\medskip

\begin{center}
	FIGURES
\end{center}

\begin{references}
\bibitem{Hegger2000}  H. Hegger, C. Petrovic, E. G. Moshopoulou, M. F.
Hundley, J. L. Sarrao, Z. Fisk, and J. D. Thompson. {\it Phys. Rev. Lett}.,
{\bf 84}, 4986 (2000).

\bibitem{Thompson2000}  J. D. Thompson, R. Movshovich, N. J. Curro, P. C.
Hammel, M. F. Hundley, M. Jaime, P. G. Pagliuso, J. L. Sarrao, C. Petrovic,
Z. Fisk, F. Bouquet, R. A. Fisher, and N. E. Phillips, preprint, 2000.

\bibitem{Murphy}  T. P. Murphy, Donavan Hall, E. C. Palm, S. W. Tozer, Z.
Fisk, R. G. Goodrich, P.G. Pagliuso, J.L. Sarrao, J.D. Thompson, to be
published.

\bibitem{GGL-Nature}  N.~D. Mathur, F.~M. Grosche, S.~R. Julian, I.~R.
Walker, D.~M. Freye, R.~K.~W. Haselwimmer, and G.~G. Lonzarich, Nature {\bf
394}, 39 (1998).

\bibitem{Movshovich2001}  R. Movshovich, M. Jaime, J. D. Thompson, C.
Petrovic, Z. Fisk, P. G. Pagliuso, and J. L. Sarrao. preprint,
cond-mat/001135, 2001.

\bibitem{Shoenberg}  D. Shoenberg. Magnetic oscillations in metals.
(Cambridge University Press, Cambridge), 1984.

\bibitem{Wasserman}  A. Wasserman and M. Springford, {\it Adv. Phys.}, {\bf %
45}, 471, (1996).

\bibitem{Haga2000}  Y. Haga, Y. Inada, H. Harima, K. Oikawa, M. Murakawa, H.
Nakawaki, Y. Tokiwa, D. Aoki, H. Shishido, S. Ikeda, N. Watanabe, and Y.
Onuki, preprint (2000).

\bibitem{Hall}  Donavan Hall, T. Murphy, E. Palm, S. Tozer, Z. Fisk, R. G.
Goodrich, P. G. Pagliuso and J. L. Sarrao, {\it Phys. Rev}. B, to be
published. cond-mat/0011395

\bibitem{Petrovic2000}  C. Petrovic, P. G. Pagliuso, M. F. Hundley, R.
Movschovich, J. L. Sarrao, J. D. Thompson, and Z. Fisk. preprint (2000).

\bibitem{Petrovic2001}  C. Petrovic, R. Movshovic, M. Jaime, P. G. Pagliuso,
M. F. Hundley, J. L. Sarrao, Z. Fisk, and J. D. Thompson. preprint (2001)
cond-mat/0012261
\end{references}
\end{document}